\documentclass[10pt]{iopart}

\usepackage[pdftex,breaklinks=true,bookmarksnumbered]{hyperref}

\usepackage{graphics,graphicx}
\usepackage{psfrag}
\usepackage{dcolumn}
\usepackage{psfrag}
\usepackage{iopams}
\usepackage{amssymb}
\usepackage{amsfonts}
\usepackage{bbm}
\usepackage{amscd}
\usepackage{array}
\usepackage{tensind}
\usepackage{mathrsfs}
\tensordelimiter{?}

\usepackage{color}

\newcommand{\be}{\begin{equation}}
\newcommand{\ee}{\end{equation}}
\newcommand{\bea}{\begin{eqnarray}}
\newcommand{\eea}{\end{eqnarray}}

\def\bes{\begin{subequations}}
\def\ens{\end{subequations}}
\def\bal{\begin{align}}
\def\eal{\end{align}}

\def\vct#1{\mathbf{#1}}

\def\user#1#2{32}

\newcommand{\pa}{\partial}

\newcommand{\text}[1]{\mathrm{#1}}

\def\pipii{(\mathbf{p}_1 \cdot \mathbf{p}_2)}

\def\Sin{(\mathbf{S}_1 \cdot \mathbf{n}_{1 2})}

\def\pipi{\mathbf{p}_1^2}
\def\piipii{\mathbf{p}_2^2}

\def\SiSi{\mathbf{S}_1^2}

\def\SipiNW{(\hat{\mathbf{S}}_1 \cdot \mathbf{p}_1)}

\def\SipiiNW{(\hat{\mathbf{S}}_1 \cdot \mathbf{p}_2)}

\def\SinNW{(\hat{\mathbf{S}}_1 \cdot \hat{\mathbf{n}}_{1 2})}

\def\pinNW{(\mathbf{p}_1 \cdot \hat{\mathbf{n}}_{1 2})}
\def\piinNW{(\mathbf{p}_2 \cdot \hat{\mathbf{n}}_{1 2})}

\def\SiSiNW{\hat{\mathbf{S}}_1^2}

\def\nl{\nonumber \\ &&}

\begin{document}

\title[NLO Spin-Squared Hamiltonian for Compact Binaries]
	{Reduced Hamiltonian for next-to-leading order Spin-Squared
        Dynamics of General Compact Binaries}
\author{Steven Hergt\footnote[1]{Email: steven.hergt@uni-jena.de}, Jan Steinhoff\footnote[2]{Email: jan.steinhoff@uni-jena.de} and Gerhard Sch\"afer\footnote[3]{Email: g.schaefer@tpi.uni-jena.de}}
\address{Theoretisch--Physikalisches Institut, Friedrich--Schiller--Universit\"at, Max--Wien--Platz 1, 07743 Jena, Germany, EU}
\date{\today}

\begin{abstract}
Within the post Newtonian framework the fully reduced Hamiltonian 
(i.e., with eliminated spin supplementary condition) for the next-to-leading order
spin-squared dynamics of general compact binaries is presented. The Hamiltonian is applicable to the spin dynamics of all
kinds of binaries with self-gravitating components like black holes and/or neutron
stars taking into account spin-induced quadrupolar deformation effects in second
post-Newtonian order perturbation theory of Einstein's field equations.
The corresponding equations of motion for spin, position and momentum variables
are given in terms of canonical Poisson brackets. Comparison with a nonreduced
potential calculated within the Effective Field Theory approach is made.
\end{abstract}
\maketitle

\section{Introduction}
A crucial prediction of Einstein's theory of General Relativity is the
existence of gravitational waves (GWs), e.g. resulting from the
inspiralling and merging process of two compact objects. Up till now
those waves are purely theoretical predictions with lack of direct
experimental verification, but their direct detection is under
preparation by gravitational wave observatories on Earth, e.g., LIGO, VIRGO,
GEO, or LISA, a future space mission \cite{Rowan:Hough:2000}.

A first indirect evidence for the existence of GWs was the
observation of energy loss in the orbital motion of the Hulse-Taylor
binary pulsar PSR B1913+16 being in full agreement with the predictions of Einstein's
theory using the quadrupole radiation formula. This discovery was awarded
the Nobel Prize in 1993. In the meantime another strong indirect
evidence occurred with the double pulsar system PSR J0737-3039A and B
\cite{Kramer:Stairs:Manchester:McLaughlin:Lyne:others:2006,Kramer:Wex:2009}.
For analysis of the measured GW patterns one has to provide very accurate
templates following from theory. This can be achieved by numerical
calculations with the matching of functions to the results or
directly using analytic tools. In the latter case, waveforms can be
obtained only perturbatively due to missing analytic solutions of
the Einstein field equations for two or more (spinning) compact objects (black
holes (BHs), neutron stars (NHs)). One of the most successful approximation methods
is the post-Newtonian one in which the metric keeps close to the flat
spacetime relying on the assumption that the typical velocity $v$
in a system divided by the speed of light $c$ is always small,
$v/c\sim\epsilon \ll 1$.
The deviations from the flat metric can be characterized by the Newtonian
potential $\Phi$; so for a binary system, $\Phi/c^2\sim v^2/c^2\sim\epsilon^2$.
In an appropriate limit (as
$\epsilon\rightarrow 0$), the post-Newtonian (PN) approximation yields Newton's
equations.

The merging process of two compact objects is divided into four
time scale sectors: inspiral, plunge, merger, and ring-down. Each sector delivers
characteristic theoretical wave patterns, that are hoped to be matched against
measured signals in the future. The PN approximation provides an excellent
analytic handling for the inspiral phase. If the PN calculations are very accurate
and thus high in order one can make very sensible predictions when
comparing with measured signals. In this article we focus on calculations of the
next-to-leading order (NLO) dynamics of spin-induced quadrupolar deformation
effects. Surely, the most compact dynamical object is the Hamiltonian,
generating the equations of motion, so we calculate in section \ref{ADM} the NLO
spin-squared one including a constant $C_Q$ parameterizing spin-induced
quadrupolar deformation effects. $C_Q$ can be given definite values describing
black holes (BHs) or neutron stars (NSs). For neutron stars $C_Q$ also depends
on the model or equation of state (EoS). Thus our result, as it seems to be 
necessary to accurately measure $C_Q$ within future GW astronomy, could help to
find the right EoS. The Hamiltonian in the present paper is calculated within the
canonical formalism of Arnowitt, Deser, and Misner (ADM)
\cite{Arnowitt:Deser:Misner:1962}. It should be noted that our Hamiltonian is
fully reduced in the sense that the spin supplementary condition (SSC)
is  eliminated on the level of the Hamiltonian. Further we make a formal counting of
the spin as $c^0$ and do not distinguish between fast and slowly spinning objects
(see, e.g., \cite{Hergt:Schafer:2008} and also Appendix A of
\cite{Steinhoff:Wang:2009}).

By now there are a lot of results regarding spin effects at the conservative
orders in the PN approximation. The leading order (LO) spin effects are
well-known for black holes, see, e.g., \cite{Barker:OConnell:1975,DEath:1975,
Barker:OConnell:1979,Thorne:Hartle:1985,Thorne:1980}. The LO $C_Q$-dependence is given in
\cite{Barker:OConnell:1979,Poisson:1998}. The NLO spin effects were only tackled recently. The first
derivation of the NLO spin-orbit (SO) equations of motion (EoM) is given in 
\cite{Tagoshi:Ohashi:Owen:2001} which became further developed in
\cite{Faye:Blanchet:Buonanno:2006}, both in harmonic gauge. Later, 
within the ADM canonical formalism, a Hamiltonian presentation was achieved \cite{Damour:Jaranowski:Schafer:2008:1}
(see also \cite{Steinhoff:Schafer:Hergt:2008}).
The NLO spin(1)-spin(2) dynamics was found
in \cite{Steinhoff:Hergt:Schafer:2008:2,Steinhoff:Schafer:Hergt:2008} and
confirmed by \cite{Porto:Rothstein:2008:1,Levi:2008}. Higher PN orders linear in
spin were tackled recently in \cite{Steinhoff:Schafer:2009:2,Barausse:Racine:Buonanno:2009,
Steinhoff:Wang:2009}. In particular, Ref.\ \cite{Steinhoff:Schafer:2009:2}
extended the point-mass ADM formalism to spinning objects, valid to any order
linear in spin. Even Hamiltonians of cubic and higher order in spin were
obtained for binary black holes (BBHs) 
\cite{Hergt:Schafer:2008:2,Hergt:Schafer:2008}. Besides quadrupolar
deformations induced by proper rotation (spin) and treated in the present
paper, tidal deformations induced through the gravitational field of
the other object were also treated, see, e.g., \cite{Hartle:1974,Damour:Nagar:2009,
Taylor:Poisson:2008}.

A nonreduced potential (i.e., with SSC not eliminated on the level of the
potential) corresponding to the result of the present paper was already
calculated in \cite{Porto:Rothstein:2008:2}, however, a term relevant for the
center-of-mass motion was missing and only found recently
\cite{Porto:Rothstein:2008:2:err}. A comparison with the result of the present
paper is not trivial, if one wants to avoid comparing all (rather long)
equations of motion; instead it is more efficient to stay on the level of the (relatively short)
potential. In section \ref{PRcompare} we sketch how to transform the
potential from \cite{Porto:Rothstein:2008:2,Porto:Rothstein:2008:2:err} into a
reduced Hamiltonian where we will find full agreement with our result of the present
paper. Considering the special case of black holes (or $C_Q = 1$), we already
succeeded in calculating the Hamiltonian of the present paper in
\cite{Steinhoff:Hergt:Schafer:2008:1,Hergt:Schafer:2008}, providing for the
first time both the spin and correct center-of-mass dynamics in this
case. There we were only able to find agreement with \cite{Porto:Rothstein:2008:2} in
the spin precession equation, see \cite{Steinhoff:Schafer:2009:1} (after
identifying a sign typo in \cite{Porto:Rothstein:2008:2}). With the correction
in \cite{Porto:Rothstein:2008:2:err} a full comparison can now be
attempted. It will be provided in section \ref{PRcompare}.

More work needs to be done for an application of the result of the present paper
to GW astronomy. In order to obtain the NLO radiation field (for the SO case see
\cite{Blanchet:Buonanno:Faye:2006,Blanchet:Buonanno:Faye:2006:err})
the stress-energy tensor has to include spin-squared corrections. This stress-energy
tensor arises from the one with a general quadrupole
\cite{Steinhoff:Puetzfeld:2009} by a spin-squared ansatz for the
mass-quadrupole, see \cite{Steinhoff:Hergt:Schafer:2008:1}.
Moreover, the NLO spin contribution should be of importance for data analysis. In a
recent publication \cite{Reisswig:Husa:Rezzolla:Dorband:Pollney:Seiler:2009} it 
has been shown that for maximal spins (aligned with the total orbital
angular momentum), the event rates are roughly thirty times larger than of those
matter systems with anti-aligned spins to orbital angular momentum and eight times
as large as for non-spinning binaries. So especially considering such sources
the event rate will increase with the inclusion of spin effects.
Further, for the creation of templates, it is useful to find a parametrization
of the orbits by solving the EoM. It is common to describe the conservative
dynamics in terms of certain orbital elements. Spin precession and dissipative
effects can then be formulated as secular EoM of the orbital elements. For
explicit solutions including spin at LO SO see, e.g.,
\cite{Konigsdorffer:Gopakumar:2005,Tessmer:2009}.

\section{The NLO Spin-Squared Hamiltonian\label{ADM}} 

We start with giving a short overview concerning the calculation of the Hamiltonian in 
question. This calculation is done within the ADM canonical formalism
\cite{Arnowitt:Deser:Misner:1962}. We use units in which $16\pi G=c=1$, where
$G$ is the Newtonian gravitational constant and $c$ the velocity of light. Greek
indices will run over $0,1,2,3$, Latin over $1,2,3$. For the signature of 
spacetime we choose +2. We employ the following notations:
${\bf x}=\left(x^i\right)$ ($i=1,2,3$) denotes a point in the 3-dimensional
Euclidean space $\mathbb{R}^3$ endowed with a standard Euclidean metric and a
scalar product (denoted by a dot).  Letters $a$ and $b$ are body labels (usually
they are set to $1$ or $2$),
so ${\bf x}_a\in\mathbb{R}^3$ denotes the position of the $a$th point mass.
We also define  ${\bf r}_a := {\bf x} - {\bf x}_a$, $r_a := |{\bf r}_a|$,
${\bf n}_a := {\bf r}_a/r_a$; and for $a\ne b$,  ${\bf r}_{ab} := {\bf x}_a -
{\bf x}_b$, $r_{ab} := |{\bf r}_{ab}|$, ${\bf n}_{ab} := {\bf r}_{ab}/r_{ab}$;
$|\cdot|$ stands here for the length of a vector.  The linear momentum vector of
the $a$th body is denoted by ${\bf p}_a=\left(p_{ai}\right)$, and $m_a$
denotes its mass parameter. The usual flat space spin vector of the $a$th body
(in local coordinates)
is denoted by $\vct{S}_a=\left(S_{a(i)}\right)$ in correspondence with our paper
\cite{Steinhoff:Hergt:Schafer:2008:1} and its associated antisymmetric tensor by 
$S_{a(i)(j)}=\epsilon_{ijk}S_{a(k)}$ with the total antisymmetric
$\epsilon$-symbol defined as $\epsilon_{123}=1$. We abbreviate 
$\delta\left({\bf x}-{\bf x}_a\right)$ by $\delta_a$. The partial
differentiation with respect to $x^i$ is denoted by $\pa_i$ or by a comma, i.e.,
$\pa_i\phi\equiv\phi_{,i}$; the partial differentiation with respect to $x_a^i$
we denote by $\pa_{ai}$. 

Following the ADM canonical formalism, the independent degress of freedom of
the gravitational field are described
by $h_{ij}^{TT}$, the transverse-traceless part of $h_{ij}=g_{ij}-\delta_{ij}$
($h^{TT}_{ii}=0$, $h^{TT}_{ij,j}=0$), and by conjugate
momenta ${\pi}^{ij\,TT}$. The needed energy and linear momentum density
expressions are given by 

\begin{eqnarray}
{\gamma}^\frac{1}{2} T^{\mu\nu}n_\mu n_\nu&=&{\cal{H}}^{\rm m(atter)}\,,\\
-{\gamma}^\frac{1}{2}T^{\;\mu}_i n_\mu&=&{\cal{H}}^{\rm m(atter)}_i\,,
\end{eqnarray}
where $\gamma=\det(g_{ij})$, $\gamma^{ij}$ is inverse to $g_{ij}$, $n^{\nu}$ 
is a unit timelike normal to hypersurface $x^0=\rm const$, and
$T^{\mu\nu}$ is the stress-momentum tensor of the matter system.
Hereafter, we call its constituents the ``particles'', but they
may well represent neutron stars or black holes.
This is substantiated by ``general relativity's adherence to the strong 
equivalence principle'': black holes and other compact bodies, to some
approximation, obey the same laws of motion as test bodies; see, e.g., \cite{DEath:1996}. Also, the analysis of the 
initial-value solutions for black holes shows that as in electromagnetism, where
image charges are described by delta functions, black holes in full general 
relativity can be represented by ``image masses'' with delta functions support 
\cite{Jaranowski:Schafer:2000:2}. It is convenient to choose the following 
four coordinate conditions 
\begin{equation}
\pi^{ii}=0\,,\quad g_{ij} = \psi^4 \delta_{ij}+h_{ij}^{TT}\,\,,\quad \psi=\left(1+\frac{1}{8}\phi \right)\,.
\end{equation} 
The standard ADM Hamiltonian (cf. \cite{Arnowitt:Deser:Misner:1962})
\begin{equation}
H = \oint \rmd S_i (g_{ij,i}-g_{jj,i}),
\end{equation} 
then becomes, using the Gauss theorem,
\begin{equation}
H=- \int \rmd^3x \Delta \phi\,.
\end{equation} 
The integrand $\Delta \phi=\partial_i\partial_i \phi$ 
can be expressed in terms of $\vct{x}_a$, $\vct{p}_a$, $\vct{S}_a$, $h^{TT}_{ij}$ 
and ${\pi}^{ij\,TT}$ using the constraint equations. By expansions of the 
field equations in powers of $G$ and after adopting suitable regularization 
procedures of integrals involved (see, e.g., Ref.\ \cite{Jaranowski:1997} and the
Appendix in \cite{Jaranowski:Schafer:1998}), one can determine the Hamiltonian.

Following the procedure outlined in our previous 
papers \cite{Hergt:Schafer:2008,Steinhoff:Hergt:Schafer:2008:1}, the Hamiltonian 
and the other generators are constructed as to fulfill the Poincar\'e algebra up to 2PN order
depending on standard canonical variables
\begin{eqnarray}
\{ x_a^{i}, p_{bj} \} &= \delta^i_j\delta_{ab} \,, \\
\{ S_a^{(i)}, S_a^{(j)} \}&= \epsilon_{ijk} S_a^{(k)} \,,
\end{eqnarray}
with all other brackets being zero. The coefficient equations resulting from this
procedure will change due to a modified Hamiltonian and CoM vector entering
the crucial relation $\{G_{i},H\}=P_{i}$ see Equation (2.4) in 
\cite{Hergt:Schafer:2008}. The modification leading to 
spin quadrupolar deformation effects of a general compact object has to be made 
in the leading order spin-squared Hamiltonian labeled as $H_{S_{1}^2}$ in equation
(2.8) in \cite{Hergt:Schafer:2008}, which now has to include a general 
spin-quadrupole constant $C_{Q}$ see \cite{Poisson:1998}; additionally an 
appropriate $S_{1}^2$-CoM vector has 
to be found. Both can be accomplished by adopting the static source expression for 
$\mathcal{H}^{m}$ from our paper \cite{Steinhoff:Hergt:Schafer:2008:1} equation (4)
and incorporating the $C_{Q}$ constant reading
\begin{eqnarray}\label{staticsource}
	\mathcal{H}^{\rm matter}_{S_1^2, \rm static} =&
		\frac{c_1}{m_1} \left( I^{ij}_1 \delta_1 \right)_{; ij}+ \frac{1}{8 m_1} g_{mn} \gamma^{pj} \gamma^{ql} \gamma^{mi}_{~~,p} \gamma^{nk}_{~~,q} \hat{S}_{1 ij} \hat{S}_{1 kl} \delta_1 \nonumber\\
		&+ \frac{1}{4m_1} \left( \gamma^{ij} \gamma^{mn} \gamma^{kl}_{~~,m} \hat{S}_{1 ln} \hat{S}_{1 jk} \delta_1 \right)_{,i} \,,\\
                                               \quad\qquad c_{1}=&-\frac{1}{2}C_{Q}\,,
\end{eqnarray}
meaning $C_{Q}=1$ for BH. Symbolic abbreviations in this formula are
taken unaltered from the original paper thus denoting the same
mathematical objects. This means that $S_{(i)}$ being given
in an Euclidean basis can be related to a spin tensor $\hat{S}_{ij}$
in a coordinate basis with the help of a triad (dreibein) $e_{i(j)}$
by $\hat{S}_{ij} = e_{i(k)} e_{j(l)} \epsilon_{klm} S_{(m)}$.
The dreibein as a function of the metric is just $e_{i(j)} =\psi^2 \delta_{ij}$
because the metric can be taken as conformally flat, $g_{ij} = \psi^4 \delta_{ij}$,
in our approximation. The mass-quadrupole tensor of object 1, $I_1^{ij}$, is given by
\begin{eqnarray}
I_1^{ij} &\equiv \gamma^{ik} \gamma^{jl} \gamma^{mn} \hat{S}_{1km} \hat{S}_{1nl} + \frac{2}{3} \vct{S}^2_1 \gamma^{ij} \,, \\
2 \vct{S}_1^2 &= \gamma^{ik} \gamma^{jl} \hat{S}_{1ij} \hat{S}_{1kl} = \text{const} \,.
\end{eqnarray}
The relation to $Q^{ij}_1$ and $\vct{a}^2_1$
is $I_1^{(i)(j)} = m_1^2 Q^{ij}_1$ and $\vct{S}_1^2 = m_1^2 \vct{a}^2_1$, so in 
leading order the related quadrupole-moment tensor $Q^{ij}_1$ is just given by
\begin{equation}\label{QLO}
 Q^{ij}_1=a^{(i)}_{1}a^{(j)}_{1}-\frac{1}{3}\vct{a}^2_{1}\delta^{ij}
\end{equation}

This static source alone is also enough to determine all the $G^2$ 
terms (static, free of linear momenta) of the Hamiltonian in question. The LO 
spin-squared Hamiltonian and the 
$S_{1}^2$-CoM vector are calculated via the formulae $H=- \int \rmd^3x \Delta \phi$
and $G_{i}=-\int \rmd^3x\,x^i\Delta\phi$, respectively, with a post-Newtonian 
perturbatively expanded
$\Delta\phi$ and $\mathcal{H}^{\rm matter}_{S_1^2, \rm static}$ according to equations 
(4.14)\,-\,(4.16) in \cite{Hergt:Schafer:2008}. The results are
\begin{equation}
 H_{S_{1}^2}^{C_{Q}}=\frac{G}{2}\frac{m_{1}m_{2}}{r_{12}^3}C_{Q}\left(3\frac{\Sin^2}{m_{1}^2}-\frac{\SiSi}{m_{1}^2}\right)\,,
\end{equation}
and
\begin{eqnarray}
 \fl\vct{G}_{S_{1}^2}=G\frac{m_{2}}{m_{1}}\left[\nu_{1}\frac{\left(\vct{S}_{1}\cdot\vct{n}_{12}\right)\vct{S}_1}{r_{12}^2}+\frac{\left(\vct{S}_{1}\cdot\vct{n}_{12}\right)^2}{r_{12}^3}\left(\nu_{2}\vct{x}_{1}+\nu_{3}\vct{x}_{2}\right)+\frac{\vct{S}_{1}^2}{r_{12}^3}\left(\nu_{4}\vct{x}_{1}+\nu_{5}\vct{x}_{2}\right)\right]
\end{eqnarray}
with coefficients
\begin{eqnarray}
 \fl \nu_{1}&=& -\frac{1}{2}-\frac{3}{2}C_{Q}\,,\quad\nu_{2}=\frac{3}{4}C_{Q}\,,\quad\nu_{3}= \frac{3}{4}C_{Q}\,,\quad\nu_{4}= \frac{1}{2}+\frac{1}{4}C_{Q}\,,\nonumber\\
 \fl\nu_{5}&=& -\frac{1}{2}-\frac{3}{4}C_{Q}\,.
\end{eqnarray}
The non-static (momenta based) terms of the Hamiltonian will be determined
via the same ansatzes for the source terms in $\mathcal{H}^{\rm m}$ and 
$\mathcal{H}^{\rm m}_{i}$ as in \cite{Hergt:Schafer:2008} with the LO quadrupole moment
(\ref{QLO}) reading
\begin{eqnarray}
\fl \mathcal{H}^{\rm m}=&\sum_{b=1}^{2}\Bigg[-\frac{m_{b}}{2}C_{Q} Q_{b}^{ij}\partial_{i}\partial_{j}-\frac{1}{2}\vct{p}_{b}\cdot\left(\vct{a}_{b}\times\vct{\partial}\right)+\left(\gamma^{ij}p_{bi}p_{bj}+m_{b}^2\right)^{1/2}
  +\lambda_{1}\frac{\vct{p}_{b}^2}{2m_{b}}Q_{b}^{ij}\partial_{i}\partial_{j}\nonumber\\
\fl&+\frac{\lambda_{2}}{m_{b}}(\vct{p}_{b}\cdot\vct{\partial})Q_{b}^{ij}p_{bi}\partial_{j}+\frac{\lambda_{3}}{m_{b}}\vct{a}_{b}^2(\vct{p}_{b}\cdot\vct{\partial})^2 -\lambda_{8}\vct{p}_{b}\cdot\left(\vct{a}_{b}\times\vct{\partial}\right)Q_{b}^{ij}\partial_{i}\partial_{j}\Bigg]\delta_{b}\,,
\end{eqnarray}

\begin{eqnarray}
 \fl\mathcal{H}^{\rm m}_{i}=&-2\sum_{b=1}^{2}\Bigg[Q_{b}^{kl}\bigg(\lambda_{5}p_{bk}\partial_{l}\partial_{i}+\lambda_{6}p_{bi}\partial_{k}\partial_{l}+\lambda_{7}\left(\vct{p}_{b}\cdot\vct{\partial}\right)\delta_{li}\partial_{k}\bigg)\nonumber\\
\fl&+\lambda_{4}\vct{a}_{b}^2(\vct{p}_{b}\cdot\vct{\partial})\partial_{i}+\frac{m_{b}}{4}\left(\vct{a}_{b}\times\vct{\partial}\right)_{i}\left(1-\frac{1}{6}Q_{b}^{kl}\partial_{k}\partial_{l}\right)-\frac{1}{2}p_{bi}\Bigg]\delta_{b}\,.
\end{eqnarray}
Notice that the static term in $\mathcal{H}^{\rm m}$ involves the
$C_{Q}$ constant which follows from the expansion of (\ref{staticsource}) being the only modification of 
Eq. (4.11) in \cite{Hergt:Schafer:2008}. These sources allow the calculation of the 
NLO spin-squared Hamiltonian (with yet undetermined coefficients) leading to 
the same coefficient equations (4.50)\,-\,(4.62) in \cite{Hergt:Schafer:2008} except
for the test particle terms $\beta_{2}$ and $\beta_{4}$ which are just multiplied
by the quadrupole constant $C_{Q}$. These
equations have to be matched to the ones resulting from the requirement of 
fulfilling the Poincar\'e algebra which now include the $C_{Q}$ constant and for
that reason will slighty differ from the equations (3.8)\,-\,(3.22) in
\cite{Hergt:Schafer:2008}. The matching procedure then fixes all the coefficients
left attributing to the source term coefficients the values
\begin{eqnarray}
\lambda_{1}&=\frac{7}{4}-\frac{3}{2}C_{Q},\,\lambda_{2}=-\frac{5}{4}+\frac{3}{2}C_{Q},\,\lambda_{3}=-\frac{1}{24},\,\lambda_{4}=0,\,\\
\lambda_{5}&=\frac{1}{12}-\frac{C_{Q}}{6},\,\lambda_{6}=-\frac{1}{8}+\frac{C_{Q}}{4},\,\lambda_{7}=\frac{1}{8}\,,
\end{eqnarray}
which agrees with our results obtained in \cite{Hergt:Schafer:2008} when setting 
$C_{Q}=1$ (terms cubic in $\vct{a}$ are not of interest here and will be dropped).
In view of section \ref{PRcompare} we label from now on standard canonical
variables with a `hat' specifying its affiliation to the Newton-Wigner (NW) SSC 
except for the momentum which is chosen to be the same for NW and covariant SSC
due to the existence of canonical degrees of freedom.
The `hatted' variables are then called NW variables in the sense that they are
standard canonical, meaning
\begin{eqnarray}
\quad\{ \hat{x}_a^{i}, \hat{p}_{bj} \} &= \delta^i_j\delta_{ab} \,, \\
\{ \hat{S}_a^{(i)}, \hat{S}_a^{(j)} \}&= \epsilon_{ijk} \hat{S}_a^{(k)} \,,\\
\quad\qquad\hat{p}_{ai}&=p_{ai}\,,
\end{eqnarray}
all other brackets being zero. Subtleties arising from that definition of NW
variables are discussed in detail in our Comment \cite{Steinhoff:Schafer:2009:1}. 
The resulting NLO spin-squared Hamiltonian for general compact binaries reads

\begin{eqnarray}
\fl H^{\text{ADM can}}_{\text{NLO}~S_{1}^2}=&~\frac{G}{\hat{r}_{12}^3}\Bigg[\frac{m_{2}}{m_{1}^3}\Bigg(\left(-\frac{21}{8}+\frac{9}{4}C_{Q}\right)\pipi\SinNW^2\nonumber\\
\fl&+\left(\frac{15}{4}-\frac{9}{2}C_{Q}\right)\pinNW\SinNW\SipiNW+\left(-\frac{5}{4}+\frac{3}{2}C_{Q}\right)\SipiNW^2\nonumber\\
\fl&+\left(-\frac{9}{8}+\frac{3}{2}C_{Q}\right)\pinNW^2\SiSiNW+\left(\frac{5}{4}-\frac{5}{4}C_{Q}\right)\pipi\SiSiNW\Bigg)\nonumber\\
              \fl&+\frac{1}{m_{1}^2}\Bigg(-\frac{15}{4}C_{Q}~\pinNW\piinNW\SinNW^2\nonumber\\
\fl&\quad+\left(3-\frac{21}{4}C_{Q}\right)\pipii\SinNW^2\nonumber\\
              \fl&\quad+\left(-\frac{3}{2}+\frac{9}{2}C_{Q}\right)\piinNW\SinNW\SipiNW\nonumber\\
\fl&\quad+\left(-3+\frac{3}{2}C_{Q}\right)\pinNW\SinNW\SipiiNW\nonumber\\
              \fl&\quad+\left(\frac{3}{2}-\frac{3}{2}C_{Q}\right)\SipiNW\SipiiNW\nonumber\\
\fl&\quad+\left(\frac{3}{2}-\frac{3}{4}C_{Q}\right)\pinNW\piinNW\SiSiNW\nonumber\\
              \fl&+\left(-\frac{3}{2}+\frac{9}{4}C_{Q}\right)\pipii\SiSiNW\Bigg)+\frac{C_{Q}}{m_{1}m_{2}}\Big(\frac{9}{4}\piipii\SinNW^2-\frac{3}{4}\piipii\SiSiNW\Big)\Bigg]\nonumber\\
              \fl&~+\frac{G^2m_{2}}{\hat{r}_{12}^4}\Bigg[\left(2+\frac{1}{2}C_{Q}+\frac{m_{2}}{m_{1}}\Big(1+2C_{Q}\Big)\right)\SiSiNW\nonumber\\
\fl&+\left(-3-\frac{3}{2}C_{Q}-\frac{m_{2}}{m_{1}}\Big(1+6C_{Q}\Big)\right)\SinNW^2\Bigg]\,,
\end{eqnarray}
for $C_Q=1$ being in full agreement with the result for BH presented for the first
time in \cite{Steinhoff:Hergt:Schafer:2008:1}.

\section{Comparison with the NLO spin(1)spin(1) potential\label{PRcompare}}
In order to transform the Routhian $R$ obtained within the Effective Field
Theory (EFT) approach 
\cite{Porto:Rothstein:2008:2,Porto:Rothstein:2008:2:err} to a nonreduced
Hamiltonian $H$, we first have to eliminate the acceleration term
\cite{Porto:Rothstein:2008:2:err} with the help of the Newtonian equations of
motion (corresponding to a redefinition of the position variables, see
\cite{Schafer:1984}). This generates correction terms to the order $G^2$ in the NLO
spin-squared potential. Next, one must replace velocities by canonical momenta
$\vct{p}_a = \frac{\partial R}{\partial \vct{v}_a}$ to get the
Hamiltonian by a Legendre transformation, i.e.,
\begin{equation}
H = \vct{v}_1 \cdot \vct{p}_1 + \vct{v}_2 \cdot \vct{p}_2 - R \,.
\end{equation}
The canonical momentum $\vct{p}_{1}$ necessary to cover all NLO spin effects
explicitly reads
\begin{eqnarray}
\vct{p}_{1} &=& \left( 1 + \frac{1}{2} \vct{v}_1^2 \right) m_1 \vct{v}_1
	+ \frac{G m_1 m_2}{2 r_{12}} [ 6 \vct{v}_1 - 7 \vct{v}_2
		- (\vct{n}_{12} \cdot \vct{v}_2) \vct{n}_{12} ] \nl
	+ \frac{G}{r_{12}^2} [ m_2 (\vct{n}_{12} \times \vct{S}_1)
		+ 2 m_1 (\vct{n}_{12} \times \vct{S}_2) ] \,, \label{canmomentum}
\end{eqnarray}
and similarly for particle 2. The Poisson brackets at this stage are
\begin{eqnarray}
\{ x^i_a, p_{bj} \} &= \delta^i_j \delta_{ab} \,, \\
\{ S^{(i)}_a, S^{(j)}_a \} &= \epsilon_{ijk} S^{(k)}_a \,, \\
\{ S^{(i)}_a, S^{(0)(j)}_a \} &= \epsilon_{ijk} S^{(0)(k)}_a \,, \\
\{ S^{(0)(i)}_a, S^{(0)(j)}_a \} &= - \epsilon_{ijk} S^{(k)}_a \,,
\end{eqnarray}
and zero otherwise.\footnote{Notice that in \cite{Porto:Rothstein:2008:1,Porto:Rothstein:2008:2} a
different sign convention was used for the Poisson brackets of the spin. As usual, we
are not showing the canonical conjugate of the spin here; see
\cite{Hanson:Regge:1974}.} Notice that these are not yet the reduced or standard
canonical brackets as $S^{(0)(i)}_a$ is still an independent degree of freedom and
was not eliminated using the covariant SSC $S^{\mu\nu}_a u_{\nu} = 0$.

It is well known that one has to proceed to Dirac brackets (DBs) if
$S^{(0)(i)}_a$ is going to be eliminated from the Hamiltonian $H$ using a SSC, see, e.g.,
\cite{Dirac:1964,Hanson:Regge:1974}. However, it is possible to find new variables
$\hat{x}_a^{i}$, $\hat{p}_{bj}$ and $\hat{S}_a^{(j)}$ for which the Dirac brackets
take on the standard form,
\begin{eqnarray}
\{ \hat{x}_a^{i}, \hat{p}_{bj} \}_{\text{DB}} &= \delta^i_j\delta_{ab} \,, \\
\{ \hat{S}_a^{(i)}, \hat{S}_a^{(j)} \}_{\text{DB}} &= \epsilon_{ijk} \hat{S}_a^{(k)} \,,
\end{eqnarray}
and zero otherwise. These new variables can only be unique up to canonical
transformations. This freedom allows us to choose $p_{1i}=\hat{p}_{1i}$,
as for the flat space case \cite{Hanson:Regge:1974}. A possible transition to
$\hat{x}_a^{i}$ and $\hat{S}_a^{j}$ then reads,
\begin{eqnarray}
 \fl S_{1(i)(j)}=&~\hat{S}_{1(i)(j)}-\bigg[\frac{p_{1[i}\hat{S}_{1(j)](k)}p_{1k}}{m_{1}^2}\left(1-\frac{3\vct{p}_{1}^2}{4m_{1}^2}\right)-\frac{2Gm_{2}}{m_{1}^2\hat{r}_{12}}p_{1[i}\hat{S}_{1(j)](k)}p_{1k}\nonumber\\
 \fl&\quad+\frac{3G}{m_{1}\hat{r}_{12}}p_{1[i}\hat{S}_{1(j)](k)}p_{1k}+\frac{G}{m_{1}\hat{r}_{12}}p_{1[i}\hat{S}_{1(j)](k)}\hat{n}_{12}^{k}(\vct{\hat{n}}_{12}\cdot\vct{p}_{2})\nonumber\\
 \fl&\quad+\frac{2Gm_{2}}{m_{1}^2\hat{r}_{12}^2}p_{1[i}\hat{S}_{1(j)](l)}\hat{S}_{1(k)(l)}\hat{n}_{12}^{k}+\frac{2G}{m_{1}\hat{r}_{12}^2}p_{1[i}\hat{S}_{1(j)](l)}\hat{S}_{2(k)(l)}\hat{n}_{12}^{k}\bigg]\,,
\end{eqnarray}
\begin{eqnarray}
 \fl x_{1}^{i}=&~\hat{x}_{1}^{i}-\bigg[\frac{1}{2m_{1}^2}p_{1k}\hat{S}_{1(i)(k)}\left(1-\frac{\vct{p}_{1}^2}{4m_{1}^2}\right)-G\frac{m_{2}}{m_{1}^2}\frac{p_{1k}\hat{S}_{1(i)(k)}}{\hat{r}_{12}}\nonumber\\
\fl&\quad+\frac{3}{2}G\frac{p_{2k}\hat{S}_{1(i)(k)}}{m_{1}\hat{r}_{12}}+\frac{G}{2}\frac{\hat{n}_{12}^{k}(\vct{\hat{n}}_{12}\cdot\vct{p}_{2})\hat{S}_{1(i)(k)}}{m_{1}\hat{r}_{12}}\nonumber\\
              \fl&\quad+G\frac{m_{2}}{m_{1}^2}\frac{\hat{S}_{1(k)(l)}\hat{S}_{1(i)(l)}\hat{n}_{12}^{k}}{\hat{r}_{12}^2}+G\frac{\hat{n}_{12}^{k}\hat{S}_{1(i)(l)}\hat{S}_{2(k)(l)}}{m_{1}\hat{r}_{12}^2}\bigg]\,,
\end{eqnarray}
where the antisymmetrization of indices pertaining to a tensor $A_{ij}$ is defined
as $A_{[ij]}=1/2(A_{ij}-A_{ji})$.
The rather complicated form of these variable transformations reflects the
complicated structure of the DBs for self-interacting spinning objects
in the covariant SSC. We will elaborate on its specific calculation in another
paper. Notice that these results are applicable to all NLO spin effects (for the
spin-orbit contributions the corrected form of the tetrad from
\cite{Steinhoff:Schafer:2009:1} has to be inserted into the SSC). To best of our
knowledge this is the first time that DBs are applied to
gravitationally self-interacting spinning objects. In
\cite{Barausse:Racine:Buonanno:2009} test-spinning objects are considered and
\cite{Hanson:Regge:1974} covers the flat space case only.

The Routhian from \cite{Porto:Rothstein:2008:2} now leads us to the reduced
NLO spin-squared Hamiltonian in the form
\begin{eqnarray}
 \fl H^{\text{EFT can}}_{\text{NLO}S_{1}^2}=&~\frac{G}{\hat{r}_{12}^3}\Bigg[\frac{m_{2}}{m_{1}^3}\Bigg(\left(-\frac{21}{8}+\frac{9}{4}C_{Q}\right)\pipi\SinNW^2\nonumber\\
\fl&+\left(\frac{21}{4}-\frac{9}{2}C_{Q}\right)\pinNW\SinNW\SipiNW+\left(-\frac{7}{4}+\frac{3}{2}C_{Q}\right)\SipiNW^2\nonumber\\
\fl&+\left(-\frac{21}{8}+\frac{9}{2}C_{Q}\right)\pinNW^2\SiSiNW+\left(\frac{7}{4}-\frac{9}{4}C_{Q}\right)\pipi\SiSiNW\Bigg)\nonumber\\
              \fl&+\frac{1}{m_{1}^2}\Bigg(-\frac{15}{4}C_{Q}~\pinNW\piinNW\SinNW^2\nonumber\\
\fl&+\left(3-\frac{21}{4}C_{Q}\right)\pipii\SinNW^2\nonumber\\
\fl&+\left(-3+\frac{9}{2}C_{Q}\right)\piinNW\SinNW\SipiNW\nonumber\\
\fl&+\left(-3+\frac{3}{2}C_{Q}\right)\pinNW\SinNW\SipiiNW\nonumber\\
              \fl&+\left(2-\frac{3}{2}C_{Q}\right)\SipiNW\SipiiNW+\left(3-\frac{15}{4}C_{Q}\right)\pinNW\piinNW\SiSiNW\nonumber\\
              \fl&+\left(-2+\frac{13}{4}C_{Q}\right)\pipii\SiSiNW\Bigg)+\frac{C_{Q}}{m_{1}m_{2}}\Big(\frac{9}{4}\piipii\SinNW^2-\frac{3}{4}\piipii\SiSiNW\Big)\Bigg]\nonumber\\
              \fl&~+\frac{G^2m_{2}}{\hat{r}_{12}^4}\Bigg[\left(2+\frac{1}{2}C_{Q}+\frac{m_{2}}{m_{1}}\Big(\frac{1}{2}+3C_{Q}\Big)\right)\SiSiNW\nonumber\\
\fl&+\left(-3-\frac{3}{2}C_{Q}-\frac{m_{2}}{m_{1}}\Big(\frac{1}{2}+6C_{Q}\Big)\right)\SinNW^2\Bigg]\,.
\end{eqnarray}
This Hamiltonian and the one calculated in the last section should differ only 
up to a canonical transformation. It should thus be possible to generate the
difference $\Delta H_{\text{NLO}S_{1}^2}=~H^{\text{EFT can}}_{\text{NLO}S_{1}^2}-H^{\text{ADM can}}_{\text{NLO}S_{1}^2}$
by a canonical transformation of the form
\begin{equation}\label{cantrafo}
 \Delta H_{\text{NLO}S_{1}^2}=\{H_{N},g^{\text{can}}_{\text{NLO}S_{1}^2}\}\,,
\end{equation}
with $H_{N}$ being the Newtonian Hamiltonian of a two-body system and $g$ being
an appropriate generator. It turns out that with the generator
\begin{eqnarray}
  \fl g^{\text{can}}_{\text{NLO}S_{1}^2}=&\frac{G}{\hat{r}_{12}^2}\frac{m_{2}}{m_{1}^2}\Bigg[\left(-\frac{1}{2}+C_{Q}\right)\pinNW\SiSiNW+\frac{1}{2}\SinNW\SipiNW\Bigg]\,,
\end{eqnarray}
equation (\ref{cantrafo}) can be fulfilled and so agreement is achieved.
This means that the Hamiltonian
$H^{\text{ADMcan}}_{\text{NLO}S_{1}^2}$ calculated with the aid of the ADM method,
in terms of invariant physical quantities, agrees with the Routhian from
above, hence there is great confidence that
$H^{\text{ADMcan}}_{\text{NLO}S_{1}^2}$ correctly describes a binary
consisting of BHs and/or NSs or other kinds of compact objects in 
post-Newtonian Einsteinian theory. The new Hamiltonian may find
immediate application in the problem of motion of orbiting binaries as
investigated and solved in e.g., \cite{Tessmer:2009,Memmesheimer:Gopakumar:Schafer:2004}.

\ack
The authors wish to thank M.\ Tessmer for useful discussions.
This work is supported by the Deutsche Forschungsgemeinschaft (DFG) through
SFB/TR7 ``Gravitational Wave Astronomy''.

\vspace{2cm}

\bibliographystyle{utphys}
\bibliography{../references}

\providecommand{\href}[2]{#2}\begingroup\raggedright\begin{thebibliography}{10}

\bibitem{Rowan:Hough:2000}
S.~Rowan and J.~Hough, ``Gravitational wave detection by interferometry (ground
  and space),'' {\em Living Rev. Relativity} {\bf 3} (2000)  3.
  \url{http://www.livingreviews.org/lrr-2000-3}.

\bibitem{Kramer:Stairs:Manchester:McLaughlin:Lyne:others:2006}
M.~Kramer, I.~H. Stairs, R.~N. Manchester, M.~A. McLaughlin, A.~G. Lyne, R.~D.
  Ferdman, M.~Burgay, D.~R. Lorimer, A.~Possenti, N.~D'Amico, J.~M. Sarkissian,
  G.~B. Hobbs, J.~E. Reynolds, P.~C.~C. Freire, and F.~Camilo, ``{T}ests of
  general relativity from timing the double pulsar",''
  \href{http://dx.doi.org/10.1126/science.1132305}{{\em Science} {\bf 314}
  (2006)  97--102},
\href{http://arxiv.org/abs/astro-ph/0609417}{{\tt arXiv:astro-ph/0609417}}.

\bibitem{Kramer:Wex:2009}
M.~Kramer and N.~Wex, ``The double pulsar system: a unique laboratory for
  gravity,'' \href{http://dx.doi.org/10.1088/0264-9381/26/7/073001}{{\em Class.
  Quant. Grav.} {\bf 26} (2009) no.~7, }.

\bibitem{Arnowitt:Deser:Misner:1962}
R.~L. Arnowitt, S.~Deser, and C.~W. Misner, ``The dynamics of general
  relativity,'' in {\em Gravitation: An Introduction to Current Research},
  L.~Witten, ed., pp.~227--265.
\newblock John Wiley, New York, 1962.
\newblock
\href{http://arxiv.org/abs/gr-qc/0405109}{{\tt arXiv:gr-qc/0405109}}.
\newblock

\bibitem{Hergt:Schafer:2008}
S.~Hergt and G.~Sch{\"a}fer, ``Higher-order-in-spin interaction {H}amiltonians
  for binary black holes from {P}oincar{\'e} invariance,''
  \href{http://dx.doi.org/10.1103/PhysRevD.78.124004}{{\em Phys. Rev. D} {\bf
  78} (2008)  124004},
\href{http://arxiv.org/abs/0809.2208}{{\tt arXiv:0809.2208 [gr-qc]}}.

\bibitem{Steinhoff:Wang:2009}
J.~Steinhoff and H.~Wang, ``Canonical formulation of gravitating spinning
  objects at 3.5 post-{N}ewtonian order,''
  \href{http://dx.doi.org/10.1103/PhysRevD.81.024022}{{\em Phys. Rev. D} {\bf
  81} (2010)  024022},
\href{http://arxiv.org/abs/0910.1008}{{\tt arXiv:0910.1008 [gr-qc]}}.

\bibitem{Barker:OConnell:1975}
B.~M. Barker and R.~F. O'Connell, ``Gravitational two-body problem with
  arbitrary masses, spins, and quadrupole moments,''
\href{http://dx.doi.org/10.1103/PhysRevD.12.329}{{\em Phys. Rev. D} {\bf 12}
  (1975)  329--335}.

\bibitem{DEath:1975}
P.~D. D'Eath, ``Interaction of two black holes in the slow-motion limit,''
  \href{http://dx.doi.org/10.1103/PhysRevD.12.2183}{{\em Phys. Rev. D} {\bf 12}
  (1975)  2183--2199}.

\bibitem{Barker:OConnell:1979}
B.~M. Barker and R.~F. O'Connell, ``The gravitational interaction: Spin,
  rotation, and quantum effects---a review,''
  \href{http://dx.doi.org/10.1007/BF00756587}{{\em Gen. Relativ. Gravit.} {\bf
  11} (1979)  149--175}.

\bibitem{Thorne:Hartle:1985}
K.~S. Thorne and J.~B. Hartle, ``Laws of motion and precession for black holes
  and other bodies,''
\href{http://dx.doi.org/10.1103/PhysRevD.31.1815}{{\em Phys. Rev. D} {\bf 31}
  (1985)  1815--1837}.

\bibitem{Thorne:1980}
K.~S. Thorne, ``Multipole expansions of gravitational radiation,''
\href{http://dx.doi.org/10.1103/RevModPhys.52.299}{{\em Rev. Mod. Phys.} {\bf
  52} (1980)  299--339}.

\bibitem{Poisson:1998}
E.~Poisson, ``Gravitational waves from inspiraling compact binaries: {T}he
  quadrupole-moment term,''
  \href{http://dx.doi.org/10.1103/PhysRevD.57.5287}{{\em Phys. Rev. D} {\bf 57}
  (1998)  5287--5290},
\href{http://arxiv.org/abs/gr-qc/9709032}{{\tt arXiv:gr-qc/9709032}}.

\bibitem{Tagoshi:Ohashi:Owen:2001}
H.~Tagoshi, A.~Ohashi, and B.~J. Owen, ``Gravitational field and equations of
  motion of spinning compact binaries to 2.5 post-{N}ewtonian order,''
  \href{http://dx.doi.org/10.1103/PhysRevD.63.044006}{{\em Phys. Rev. D} {\bf
  63} (2001)  044006},
\href{http://arxiv.org/abs/gr-qc/0010014}{{\tt arXiv:gr-qc/0010014}}.

\bibitem{Faye:Blanchet:Buonanno:2006}
G.~Faye, L.~Blanchet, and A.~Buonanno, ``{H}igher-order spin effects in the
  dynamics of compact binaries. {I}. {E}quations of motion,''
  \href{http://dx.doi.org/10.1103/PhysRevD.74.104033}{{\em Phys. Rev. D} {\bf
  74} (2006)  104033},
\href{http://arxiv.org/abs/gr-qc/0605139}{{\tt arXiv:gr-qc/0605139}}.

\bibitem{Damour:Jaranowski:Schafer:2008:1}
T.~Damour, P.~Jaranowski, and G.~Sch{\"a}fer, ``{H}amiltonian of two spinning
  compact bodies with next-to-leading order gravitational spin-orbit
  coupling,'' \href{http://dx.doi.org/10.1103/PhysRevD.77.064032}{{\em Phys.
  Rev. D} {\bf 77} (2008)  064032},
\href{http://arxiv.org/abs/0711.1048}{{\tt arXiv:0711.1048 [gr-qc]}}.

\bibitem{Steinhoff:Schafer:Hergt:2008}
J.~Steinhoff, G.~Sch{\"a}fer, and S.~Hergt, ``{ADM} canonical formalism for
  gravitating spinning objects,''
  \href{http://dx.doi.org/10.1103/PhysRevD.77.104018}{{\em Phys. Rev. D} {\bf
  77} (2008)  104018},
\href{http://arxiv.org/abs/0805.3136}{{\tt arXiv:0805.3136 [gr-qc]}}.

\bibitem{Steinhoff:Hergt:Schafer:2008:2}
J.~Steinhoff, S.~Hergt, and G.~Sch{\"a}fer, ``Next-to-leading order
  gravitational spin(1)-spin(2) dynamics in {H}amiltonian form,''
  \href{http://dx.doi.org/10.1103/PhysRevD.77.081501}{{\em Phys. Rev. D} {\bf
  77} (2008)  081501(R)},
\href{http://arxiv.org/abs/0712.1716}{{\tt arXiv:0712.1716 [gr-qc]}}.

\bibitem{Porto:Rothstein:2008:1}
R.~A. Porto and I.~Z. Rothstein, ``Spin(1)spin(2) effects in the motion of
  inspiralling compact binaries at third order in the post-{N}ewtonian
  expansion,'' \href{http://dx.doi.org/10.1103/PhysRevD.78.044012}{{\em Phys.
  Rev. D} {\bf 78} (2008)  044012},
\href{http://arxiv.org/abs/0802.0720}{{\tt arXiv:0802.0720 [gr-qc]}}.

\bibitem{Levi:2008}
M.~Levi, ``Next to leading order gravitational spin-spin coupling with
  {K}aluza-{K}lein reduction,''
\href{http://arxiv.org/abs/0802.1508}{{\tt arXiv:0802.1508 [gr-qc]}}.

\bibitem{Steinhoff:Schafer:2009:2}
J.~Steinhoff and G.~Sch{\"a}fer, ``Canonical formulation of self-gravitating
  spinning-object systems,''
  \href{http://dx.doi.org/10.1209/0295-5075/87/50004}{{\em Europhys. Lett.}
  {\bf 87} (2009)  50004},
\href{http://arxiv.org/abs/0907.1967}{{\tt arXiv:0907.1967 [gr-qc]}}.

\bibitem{Barausse:Racine:Buonanno:2009}
E.~Barausse, {\'E}.~Racine, and A.~Buonanno, ``{H}amiltonian of a spinning
  test-particle in curved spacetime,''
  \href{http://dx.doi.org/10.1103/PhysRevD.80.104025}{{\em Phys. Rev. D} {\bf
  80} (2009)  104025},
\href{http://arxiv.org/abs/0907.4745}{{\tt arXiv:0907.4745 [gr-qc]}}.

\bibitem{Hergt:Schafer:2008:2}
S.~Hergt and G.~Sch{\"a}fer, ``Higher-order-in-spin interaction {H}amiltonians
  for binary black holes from source terms of {K}err geometry in approximate
  {ADM} coordinates,'' \href{http://dx.doi.org/10.1103/PhysRevD.77.104001}{{\em
  Phys. Rev. D} {\bf 77} (2008)  104001},
\href{http://arxiv.org/abs/0712.1515}{{\tt arXiv:0712.1515 [gr-qc]}}.

\bibitem{Hartle:1974}
J.~B. Hartle, ``Tidal shapes and shifts on rotating black holes,''
\href{http://dx.doi.org/10.1103/PhysRevD.9.2749}{{\em Phys. Rev. D} {\bf 9}
  (1974)  2749--2759}.

\bibitem{Damour:Nagar:2009}
T.~Damour and A.~Nagar, ``Effective one body description of tidal effects in
  inspiralling compact binaries,''
  \href{http://dx.doi.org/10.1103/PhysRevD.81.084016}{{\em Phys. Rev. D} {\bf
  81} (2010)  084016},
\href{http://arxiv.org/abs/0911.5041}{{\tt arXiv:0911.5041 [gr-qc]}}.

\bibitem{Taylor:Poisson:2008}
S.~Taylor and E.~Poisson, ``Nonrotating black hole in a post-{N}ewtonian tidal
  environment,'' \href{http://dx.doi.org/10.1103/PhysRevD.78.084016}{{\em Phys.
  Rev. D} {\bf 78} (2008)  084016},
\href{http://arxiv.org/abs/0806.3052}{{\tt arXiv:0806.3052 [gr-qc]}}.

\bibitem{Porto:Rothstein:2008:2}
R.~A. Porto and I.~Z. Rothstein, ``Next to leading order spin(1)spin(1) effects
  in the motion of inspiralling compact binaries,''
  \href{http://dx.doi.org/10.1103/PhysRevD.78.044013}{{\em Phys. Rev. D} {\bf
  78} (2008)  044013},
\href{http://arxiv.org/abs/0804.0260}{{\tt arXiv:0804.0260 [gr-qc]}}.

\bibitem{Porto:Rothstein:2008:2:err}
R.~A. Porto and I.~Z. Rothstein, ``Erratum: {N}ext to leading order
  spin(1)spin(1) effects in the motion of inspiralling compact binaries,''
  \href{http://dx.doi.org/10.1103/PhysRevD.81.029905}{{\em Phys. Rev. D} {\bf
  81} (2010)  029905(E)}.

\bibitem{Steinhoff:Hergt:Schafer:2008:1}
J.~Steinhoff, S.~Hergt, and G.~Sch{\"a}fer, ``{S}pin-squared {H}amiltonian of
  next-to-leading order gravitational interaction,''
  \href{http://dx.doi.org/10.1103/PhysRevD.78.101503}{{\em Phys. Rev. D} {\bf
  78} (2008)  101503(R)},
\href{http://arxiv.org/abs/0809.2200}{{\tt arXiv:0809.2200 [gr-qc]}}.

\bibitem{Steinhoff:Schafer:2009:1}
J.~Steinhoff and G.~Sch{\"a}fer, ``Comment on two recent papers regarding
  next-to-leading order spin-spin effects in gravitational interaction,''
  \href{http://dx.doi.org/10.1103/PhysRevD.80.088501}{{\em Phys. Rev. D} {\bf
  80} (2009)  088501},
\href{http://arxiv.org/abs/0903.4772}{{\tt arXiv:0903.4772 [gr-qc]}}.

\bibitem{Blanchet:Buonanno:Faye:2006}
L.~Blanchet, A.~Buonanno, and G.~Faye, ``{H}igher-order spin effects in the
  dynamics of compact binaries. {II}. {R}adiation field,''
  \href{http://dx.doi.org/10.1103/PhysRevD.74.104034}{{\em Phys. Rev. D} {\bf
  74} (2006)  104034},
\href{http://arxiv.org/abs/gr-qc/0605140}{{\tt arXiv:gr-qc/0605140}}.

\bibitem{Blanchet:Buonanno:Faye:2006:err}
L.~Blanchet, A.~Buonanno, and G.~Faye, ``Erratum: {H}igher-order spin effects
  in the dynamics of compact binaries. {II}. {R}adiation field,''
  \href{http://dx.doi.org/10.1103/PhysRevD.75.049903}{{\em Phys. Rev. D} {\bf
  75} (2007)  049903(E)}.

\bibitem{Steinhoff:Puetzfeld:2009}
J.~Steinhoff and D.~Puetzfeld, ``{M}ultipolar equations of motion for extended
  test bodies in general relativity,''
  \href{http://dx.doi.org/10.1103/PhysRevD.81.044019}{{\em Phys. Rev. D} {\bf
  81} (2010)  044019},
\href{http://arxiv.org/abs/0909.3756}{{\tt arXiv:0909.3756 [gr-qc]}}.

\bibitem{Reisswig:Husa:Rezzolla:Dorband:Pollney:Seiler:2009}
C.~Reisswig, S.~Husa, L.~Rezzolla, E.~N. Dorband, D.~Pollney, and J.~Seiler,
  ``{G}ravitational-wave detectability of equal-mass black-hole binaries with
  aligned spins,'' \href{http://dx.doi.org/10.1103/PhysRevD.80.124026}{{\em
  Phys. Rev. D} {\bf 80} (2009)  124026},
  \href{http://arxiv.org/abs/0907.0462}{{\tt arXiv:0907.0462 [gr-qc]}}.

\bibitem{Konigsdorffer:Gopakumar:2005}
C.~K{\"o}nigsd{\"o}rffer and A.~Gopakumar, ``{P}ost-{N}ewtonian accurate
  parametric solution to the dynamics of spinning compact binaries in eccentric
  orbits: {T}he leading order spin-orbit interaction,''
  \href{http://dx.doi.org/10.1103/PhysRevD.71.024039}{{\em Phys. Rev. D} {\bf
  71} (2005)  024039}, \href{http://arxiv.org/abs/gr-qc/0501011}{{\tt
  arXiv:gr-qc/0501011}}.

\bibitem{Tessmer:2009}
M.~Tessmer, ``{G}ravitational waveforms from unequal-mass binaries with
  arbitrary spins under leading order spin-orbit coupling,''
  \href{http://dx.doi.org/10.1103/PhysRevD.80.124034}{{\em Phys. Rev. D} {\bf
  80} (2009)  124034},
\href{http://arxiv.org/abs/0910.5931}{{\tt arXiv:0910.5931 [gr-qc]}}.

\bibitem{DEath:1996}
P.~D. D'Eath, {\em Black holes: {G}ravitational interactions}.
\newblock Clarendon Press, Oxford, England, 1996.

\bibitem{Jaranowski:Schafer:2000:2}
P.~Jaranowski and G.~Sch{\"a}fer, ``Bare masses in time-symmetric initial-value
  solutions for two black holes,''
  \href{http://dx.doi.org/10.1103/PhysRevD.61.064008}{{\em Phys. Rev. D} {\bf
  61} (2000)  064008},
\href{http://arxiv.org/abs/gr-qc/9907025}{{\tt arXiv:gr-qc/9907025}}.

\bibitem{Jaranowski:1997}
P.~Jaranowski, ``Technicalities in the calculation of the 3rd post-{N}ewtonian
  dynamics,'' in {\em Mathematics of Gravitation, Part {II}: {G}ravitational
  Wave Detection}, A.~Kr{\'o}lak, ed., pp.~55--63.
\newblock Banach Center Publications, Vol.\ 41, Part II, Warszawa, 1997.

\bibitem{Jaranowski:Schafer:1998}
P.~Jaranowski and G.~Sch{\"a}fer, ``{T}hird post-{N}ewtonian higher order {ADM}
  {H}amilton dynamics for two-body point-mass systems,''
  \href{http://dx.doi.org/10.1103/PhysRevD.57.7274}{{\em Phys. Rev. D} {\bf 57}
  (1998)  7274--7291}, \href{http://arxiv.org/abs/gr-qc/9712075}{{\tt
  arXiv:gr-qc/9712075}}.

\bibitem{Schafer:1984}
G.~Sch{\"a}fer, ``Acceleration-dependent {L}agrangians in general relativity,''
  \href{http://dx.doi.org/10.1016/0375-9601(84)90947-2}{{\em Phys. Lett. A}
  {\bf 100} (1984)  128--129}.

\bibitem{Hanson:Regge:1974}
A.~J. Hanson and T.~Regge, ``{T}he relativistic spherical top,''
\href{http://dx.doi.org/10.1016/0003-4916(74)90046-3}{{\em Ann. Phys. (N.Y.)}
  {\bf 87} (1974)  498--566}.

\bibitem{Dirac:1964}
P.~A.~M. Dirac, {\em Lectures on Quantum Mechanics}.
\newblock Yeshiva University Press, New York, 1964.

\bibitem{Memmesheimer:Gopakumar:Schafer:2004}
R.-M. Memmesheimer, A.~Gopakumar, and G.~Sch{\"a}fer, ``{T}hird
  post-{N}ewtonian accurate generalized quasi-{K}eplerian parametrization for
  compact binaries in eccentric orbits,''
  \href{http://dx.doi.org/10.1103/PhysRevD.70.104011}{{\em Phys. Rev. D} {\bf
  70} (2004)  104011},
\href{http://arxiv.org/abs/gr-qc/0407049}{{\tt arXiv:gr-qc/0407049}}.

\end{thebibliography}\endgroup

\end{document}